\documentclass[final,3p,times,twocolumn]{elsarticle}

\usepackage{amssymb}
\usepackage{graphicx}
\usepackage{dcolumn}
\usepackage{bm}

\usepackage{amsmath}

\begin{document}


\begin{frontmatter}


\title{Kinetic modelling of quantum effects in laser-beam interaction}

\author{E.~N.~Nerush}
\ead{nerush@appl.sci-nnov.ru}
\author{I.~Yu.~Kostyukov}
\address{Institute of Applied Physics, Russian Academy of Sciences, 46 Ulyanov St.
603950 Nizhny Novgorod, Russia}


\begin{abstract}
We present the results of kinetic modelling of quantum effects in laser-beam interaction.
In the developed numerical model, electron-positron pair production by hard photons, hard
photon emission and the electromagnetic fields generated by the created charged particles are taken into account. 
Interaction of a relativistic electron beam with a strong laser pulse is analyzed.  It
is shown that the quantum effects can be important even for moderately
intense laser pulses when the number of emitted photons by single
electron is not large.  Electron-positron pair plasma production in
extremely-intense laser field via development of electromagnetic
cascades is also studied. The simulation results confirm the prediction of
strong laser field absorption in the self-generated electron-positron
plasma.  It is shown that the self-generated
electron-positron plasma can be an efficient source of energetic
gamma-quanta.
\end{abstract}


\end{frontmatter}

Electromagnetic cascades are one of the basic phenomena of
strong-field physics. As opposed to classical physics, in quantum
physics the decay of high-energy photon in strong electromagnetic
fields is possible. The photon can decay with creation of
electron-positron ($e^-e^+$) pair. If the energy of the electron and
the positron is high enough, they can emit new photons that can,
by-turn, decay with creation of new $e^-e^+$ pairs, etc.  This cascade
process is also called `electromagnetic (or QED) shower'.

Due to its importance, the cascades recently are intensively studied. For
example, electromagnetic showers that develop during the pass of
high-energy particle (electron, positron or photon) through constant
homogeneous
magnetic field are theoretically investigated in
Ref.~\cite{Akhiezer1994,Anguelov1999}.  In this case the energy for
creation of secondary particles is transferred from the initial
energy of the seed particle. In such shower the mean energy of a
particle decreases with the lapse of time.  The dynamics of
electromagnetic showers in strong fields of pulsars also attracts much
attention \cite{Timokhin2010}. Showers in inhomogeneous, varying
electromagnetic fields (fields of pulsars, for example) can be
self-sustained. This means that secondary particles in such shower are
accelerated by strong external electric field and the mean energy of a
particle in such cascade is constant or increases.

There are little experiments devoted to investigation of
self-sustained cascades. However, the recent progress in laser
technologies will allow to fill up this gap.  The laser intensity can
be very high in near future so that quantum electrodynamic (QED)
effects will be essential
\cite{Mourou2006,Bell2008,Fedotov2010,Sokolov2010,Bulanov2010}.
An electron (positron) moving in strong laser field can be accelerated up
to very high energy and can produce electromagnetic cascade. If the
laser intensity is high enough, the number of produced pairs can be so
great that self-generated electromagnetic fields of $e^-e^+$ plasma
can strongly affect the further cascade dynamics. Moreover, a large
portion of the laser energy can be absorbed by self-generated $e^-e^+$
plasma.  This effect is especially important to determine the
limitations on the intensity of high power lasers
\cite{Fedotov2010,Bulanov2010}.  The estimations of intensity threshold
for cascade development in circularly polarized standing electromagnetic wave are presented
in Ref.~\cite{Fedotov2010}.  However, in order to study the cascade
development and to verify estimations the self-consistent numerical
models is needed.  

We develop the self-consistent two-dimensional numerical model based
on particle-in-cell (PIC) and Monte Carlo (MC) methods \cite{Nerush2010}.  The model
uses the probability rates of photon emission and electron-positron
pair production  calculated in the framework of QED theory
\cite{Nikishov1964,Baier1998,Landau4}.  It exploits the strong
difference between the photon energy of laser-plasma fields and the
characteristic energy of hard
photons emitted by relativistic charged particles in electromagnetic
fields.  Making of use standard PIC technique the hard photos,
electrons and positrons are modeled as quasiparticles while the
laser and plasma field is calculated by numerically
solving of Maxwell's equations.  The model has been benchmarked to the
simulations performed by other MC codes \cite{Elkina2010}. The PIC
part of the model is two-dimensional version of the model used in
Ref.~\cite{Nerush2009}.
 
One of methods to probe QED in laboratory conditions is to study
the interaction of a relativistic electron beam with an intense laser pulse
\cite{Bamber1999,Sokolov2010b}. In the rest frame of the relativistic particle the
laser field is very intense and can be close to critical.  So at this
field strength quantum effects become important. In this case the
energy of a photon emitted by a beam electron undergoing oscillations
inside the laser pulse can be close to the electron energy.

   \begin{figure}
   \includegraphics[width=8cm]{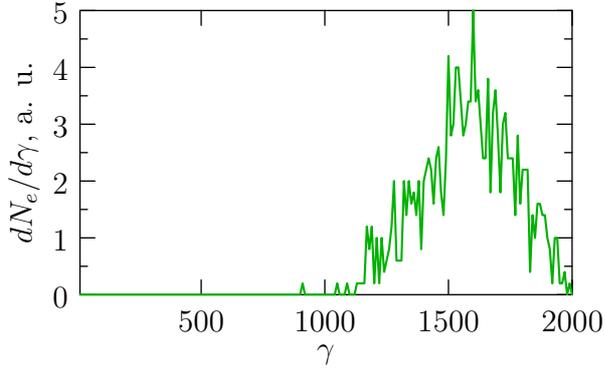}
   \caption
   {Energy spectrum of the electron bunch after its collision with a
   linearly polarized laser pulse characterised by $a_0=5$ and
   $\sigma_x=32$. The gamma-factor of the bunch electrons before the
   collision is $\gamma _0 = 2 \cdot 10^3$.}
   \label{beam1}
   \end{figure}

   \begin{figure}
   \includegraphics[width=8cm,clip]{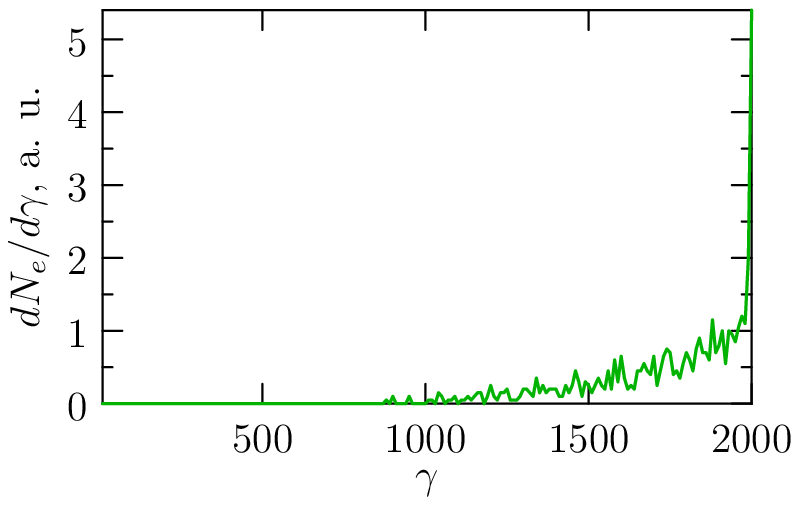}
   \caption
   {Energy spectrum of the electron bunch after its collision with a
   linearly polarized laser pulse characterised by $a_0=20$ and
   $\sigma_x=1.6$. The gamma-factor of the bunch electrons before the
   collision is $\gamma _0 = 2 \cdot 10^3$.}
   \label{beam2}
   \end{figure}

   \begin{figure}
   \includegraphics[width=8cm,clip]{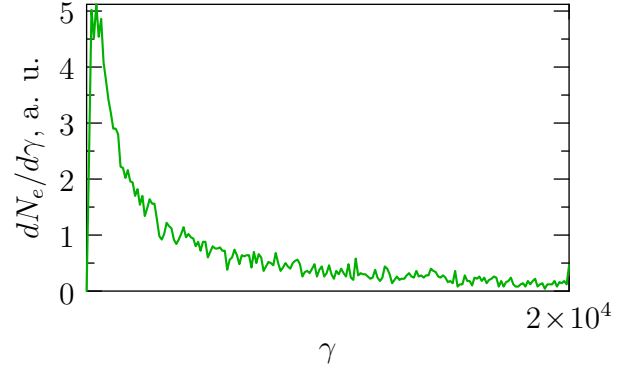}
   \caption
   {Energy spectrum of the electron bunch after its collision with a
   linearly polarized laser pulse characterised by $a_0=100$ and
   $\sigma_x=1.6$. The gamma-factor of the bunch electrons before the
   collision is $\gamma _0 = 2 \cdot 10^4$.}
   \label{beam3}
   \end{figure}

We simulate the interaction between a relativistic electron bunch and
a laser pulse by the means of our numerical model.  The laser
pulse is linearly polarized, has the Gaussian envelope $\bar E_y= \bar B_z = a_0
\exp [-y^2/\sigma_y^2-(x-t)^2 / \sigma _x^2 ]$,
where $\bar E$ and $\bar B$ are the electric and magnetic
field envelopes, respectively, $a_0 = |e|E_0/(mc\omega)$, $E_0$ is the maximal field
strength of the laser pulses, $\omega = 2\pi c/\lambda$, $\lambda$ is
the laser wavelength, $m$ and $e$ are the electron mass and charge,
respectively, $c$ is the speed of light, $\sigma_x$ is the pulse
duration, $\sigma_y$ is the pulse width. From here on coordinates are
normalized to $\lambda$ and time is normalized to $2\pi/\omega$. The
parameters of the simulation are the following. The
 wavelength
$\lambda =0.8$~$\mu $m,
$\sigma_x = 32 $, $a_0 = 5$, the laser spot
size $\sigma _y$ is assumed to be much large then the bunch width.
The initial gamma-factor of the bunch electrons is $\gamma _0 = 2000$.
For these parameters $\chi \simeq 0.03 \ll 1$ and the photon emission
regime should be classical.  The energy distribution of the electron
bunch after passing through laser pulse is shown in Fig.~\ref{beam1}.
The mean gamma-factor of the electron after interaction is $\gamma =
1568$. The task about the interaction between an electron and a laser pulse
has exact theoretical solution in the framework of Landau -- Lifshitz
representation of radiation reaction force
\cite{DiPiazza2008}.  This solution yields the following gamma-factor of the electron passing the
laser pulse with Gaussian envelope:
\begin{equation}
\gamma = \frac{\gamma _0} {1 + \mu a_0^2 \gamma _0 \sigma_x \sqrt{2 \pi^3 }},
\label{dipiazze}
\end{equation}
where $\mu = 2 e^2 \omega /(3 m c^3)$. For the given parameters this
formula yields $\gamma = 1580$ that is in good agreement with the
simulation result. 

\begin{figure}
\includegraphics[width=7cm,clip]{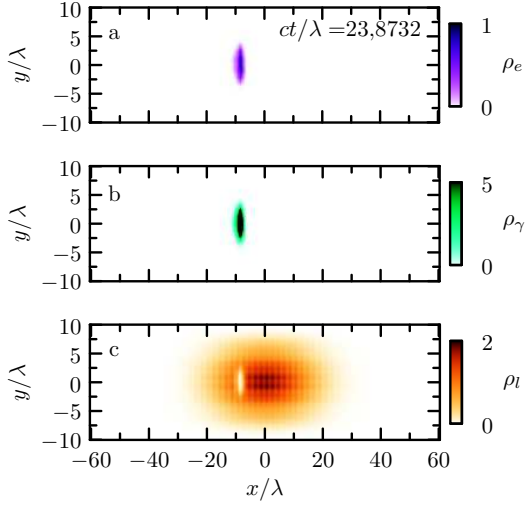}
\caption
{At the instant of time $t \approx 24$, spatial distribution of (a)
electron and (b) photon normalized densities $\rho_{e,\gamma} =
n_{e,\gamma}/(a_0 n_{cr})$ and (c) laser intensity $\rho_l =
(E^2+B^2)/E_0^2$ averaged over laser wavelength during the collision
of two laser pulses with $a_0 = 1.5 \cdot 10^3$.}
\label{movie1}
\end{figure}

\begin{figure}
\includegraphics[width=7cm,clip]{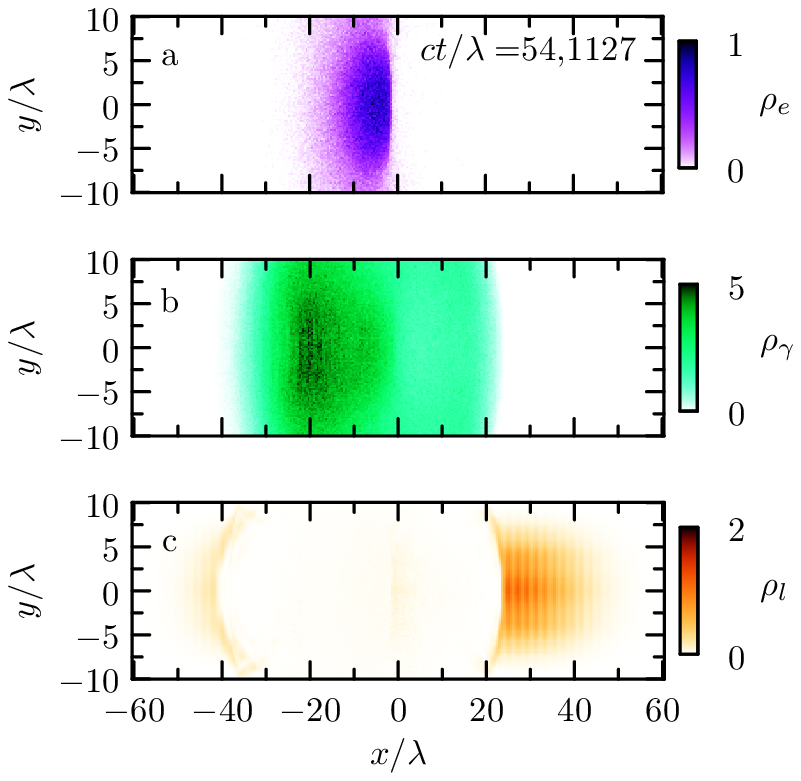}
\caption
{At the instant of time $t \approx 54$, spatial distribution of (a)
electron and (b) photon normalized densities $\rho_{e,\gamma} =
n_{e,\gamma}/(a_0 n_{cr})$ and (c) laser intensity $\rho_l =
(E^2+B^2)/E_0^2$ averaged over laser wavelength during the collision
of two laser pulses with $a_0 = 1.5 \cdot 10^3$.}
\label{movie2}
\end{figure}

\begin{figure}
\includegraphics[width=7cm,clip]{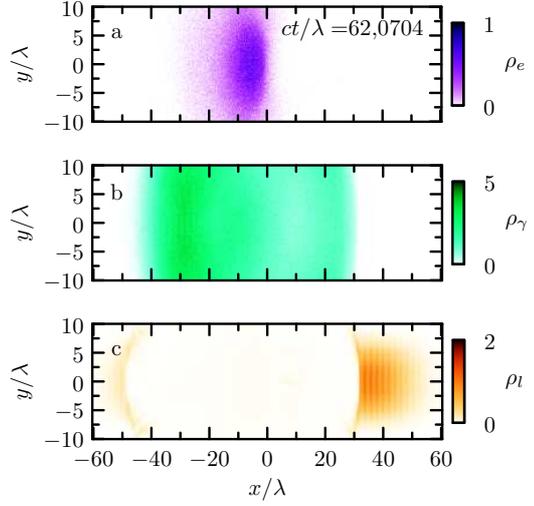}
\caption
{At the instant of time $t \approx 62$, spatial distribution of (a)
electron and (b) photon normalized densities $\rho_{e,\gamma} =
n_{e,\gamma}/(a_0 n_{cr})$ and (c) laser intensity $\rho_l =
(E^2+B^2)/E_0^2$ averaged over laser wavelength during the collision
of two laser pulses with $a_0 = 1.5 \cdot 10^3$.}
\label{movie3}
\end{figure}

It is interesting to note that quantum nature of photon emission can reveal itself 
even in classical regime ($\chi \lesssim 1$).
It is seen from Fig.~\ref{beam1} that the energy spread of the
electron bunch is large in contrast to the classical theory.  The
reason is that each bunch electron emits a small number of
photons.  The characteristic time between consecutive photon
emissions (photon emission time) can be estimated as the ratio of
the characteristic energy of emitted photon to the radiation power . In the classical limit the
photon energy is $\hbar \omega_{em} \sim a_0 \gamma^2 \hbar \omega$ and in the
quantum limit $\hbar \omega_{em} \sim \gamma m c^2$ \cite{Baier1998}.
Making of use the expression for the radiation power in the classical
and quantum limits \cite{Landau4,Baier1998} we can find the photon
emission time in the classical and quantum limits 
\begin{equation}
t_{rad} \simeq \frac{\hbar c} {e^2} \left( 1 + \chi^{1/3} \right)  t_f,
\label{trad1}
\end{equation}
where $c t_f$ is so-called radiation formation length, which is equal
to length of the electron trajectory path, over which the particle is
deflected by angle $1/\gamma $
\cite{Baier1998,Uggerhoj2005,Nerush2007}.  The radiation formation
length for an electron colliding with ultra-high intense laser pulse can be
estimated as follows: $t_f = 1/(2 \pi a_0)$.  For the given
parameters $\sigma _x /( c t_{rad}) \simeq 6$ that is not much larger
than unity.  Therefore the number of photons emitted by a bunch
electron passing through the laser pulse is not large. Because of
quantum nature of photon emission the spread in the number of
the photons and energy of the photons emitted by bunch electrons
can be significant.
 
\begin{figure}
\includegraphics[width=7cm,clip]{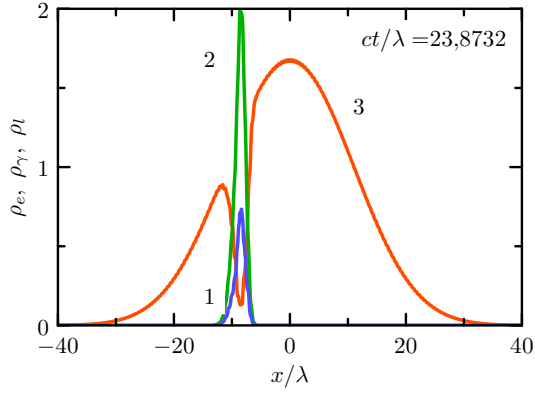}
\caption
{At the instant of time $t \approx 24$, spatial distribution
at $y=0$ of
electron normalized density $\rho_e =
n_e/(a_0 n_{cr})$ (line 1), photon normalized density $\rho_{\gamma} =
n_{\gamma}/(5 a_0 n_{cr})$ (line 2) and laser intensity $\rho_l =
(E^2+B^2)/E_0^2$ averaged over the laser wavelength (line 3) during the collision
of two laser pulses with $a_0 = 1.5 \cdot 10^3$.}
\label{nvsx1}
\end{figure}

\begin{figure}
\includegraphics[width=7cm,clip]{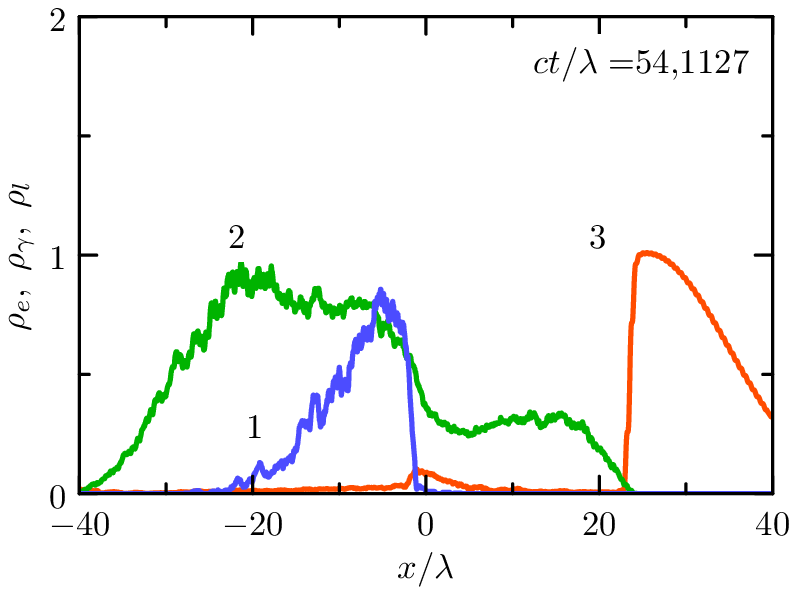}
\caption
{At the instant of time $t \approx 54$, spatial distribution
at $y=0$ of
electron normalized density $\rho_e =
n_e/(a_0 n_{cr})$ (line 1), photon normalized density $\rho_{\gamma} =
n_{\gamma}/(5 a_0 n_{cr})$ (line 2) and laser intensity $\rho_l =
(E^2+B^2)/E_0^2$ averaged over the laser wavelength (line 3) during the collision
of two laser pulses with $a_0 = 1.5 \cdot 10^3$.}
\label{nvsx2}
\end{figure}

\begin{figure}
\includegraphics[width=7cm,clip]{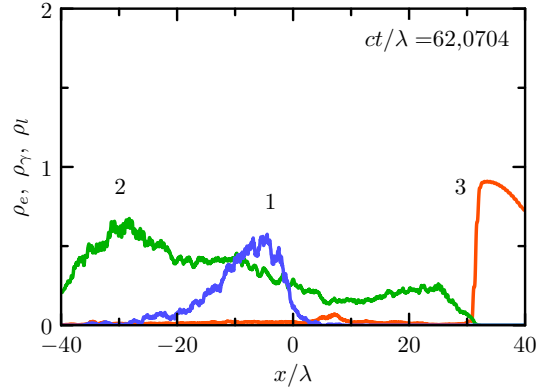}
\caption
{At the instant of time $t \approx 62$, spatial distribution
at $y=0$ of
electron normalized density $\rho_e =
n_e/(a_0 n_{cr})$ (line 1), photon normalized density $\rho_{\gamma} =
n_{\gamma}/(5 a_0 n_{cr})$ (line 2) and laser intensity $\rho_l =
(E^2+B^2)/E_0^2$ averaged over the laser wavelength (line 3) during the collision
of two laser pulses with $a_0 = 1.5 \cdot 10^3$.}
\label{nvsx3}
\end{figure}

The electron energy distribution of the electron bunch after
interaction with shorter and more intense laser pulse ($\sigma_x = 1.6
$, $a_0 = 20$) is shown in Fig.~\ref{beam2}. The initial
electron gamma-factor is $\gamma_0=2000$. For these parameters $\chi
\simeq 0.1 $  and $t_{rad} \simeq 1.1 $.  As $\sigma _x / 
t_{rad} \simeq 1$ a bunch electron emits approximately one
photon in the average.  As the electron distribution function peaks at $\gamma =
2000$, a significant number of the electrons emit no photons.  It
follows from the obtained results presented in Figs.~\ref{beam1} and
\ref{beam2} that even in the limit $\chi \ll 1$ when the regime of the
photon emission is classical and the total loss of the bunch energy
due to the photon emission can be calculated in the classical approach
the distribution function of the electron bunch after interaction
cannot be described in the framework of classical electrodynamics if
the mean photon number emitted by the bunch electron is small.

We model the interaction between electron bunch and the laser pulse in
quantum regime. The parameters of the laser pulse and the electron
bunch are  $\sigma_x = 1.6 $, $a_0 = 100$ and $\gamma _0 = 2
\times 10^4$. For these parameters $\chi \simeq 6 $  and $
t_{rad} \simeq 0.2$. The electron distribution function after
interaction with the laser pulse is shown in Fig.~\ref{beam3}. It is
seen from Fig.~\ref{beam1} that the bunch lost most of the initial
energy.  Moreover, some of the emitted photons decay and produce
electron-positron pair.  The similar effects are observed in the
experiments \cite{Bamber1999} and discussed, for example, in
Ref.~\cite{Sokolov2010b}.

In this work we also present the results of 2D numerical simulations
of QED cascade in the field of two colliding linearly polarized laser
pulses. Laser pulses have the Gaussian envelopes at initial instant
$t=0$  and propagate along $x$-axis towards each other. At initial
instant of time the distance between laser pulses is equal to
$2\sigma_x$ and coordinates of the pulse centers are $x = -\sigma_x$ and
$x = \sigma_x$. The cascade was initiated by $5 \cdot 10^8$ $1
\text{~GeV}$ photons situated near $x=-\sigma_x$, $y=0$. The values of
simulation parameters are the following: $\lambda =
1.24\text{~}\mu\text{m}$, $a_0 = 1.5 \cdot 10^3$, $\sigma_x = 19$,
$\sigma_y = 8$.

The spatial distributions of normalized electron and photon densities
and laser intensity at three instants of time are shown at
Figs.~\ref{movie1}, \ref{movie2} and \ref{movie3}. At initial stage of
cascade development (Fig.~\ref{movie1}) $e^-e^+$ plasma density
rapidly
reaches the value of the relativistic critical density $a_0 n_{cr}$, where
$n_{cr} = m\omega^2/ ( 8 \pi e^2 )$ is the nonrelativistic critical
density for $e^-e^+$ plasma. The $x$-scale of created plasma
is about the laser wavelength. Creation of dense plasma leads to
substantial absorption of laser pulses and to the decrease of laser
fields in the region occupied by the plasma. At later stages of the
cascade development (Figs.~\ref{movie2}, \ref{movie3}) $e^-e^+$ plasma expands.

Strong absorption of the laser pulses starts at $t \approx 24$ when the plasma density becomes
close to $a_0 n_{cr}$.  Due to the asymmetry in the initial
position of the seed particles one of the
laser pulses partially has passed the region occupied by the plasma at $t \approx 24$,
while only small part of the front side of the other pulse has passed
this region. Then the laser pulse that starts at $x=\sigma_x$ and the
rear part of the other laser pulse are mostly absorbed by created
$e^-e^+$ plasma (see Fig.~\ref{movie2}~(c)).  The asymmetry in the
initial conditions also leads to asymmetry in the distribution of hard
photons emitted by the electrons and the positrons. Normalized electron
and photon densities and the distribution of laser intensity averaged
over the laser wavelength at $x$-axis are shown on
Figs.~\ref{nvsx1}-\ref{nvsx3}. The asymmetry of the laser pulses
absorption and hard photon emission can be also seen in
Figs.~\ref{nvsx1}-\ref{nvsx3}. The steeping of the right edge of
the electron distribution caused by momentum transmission from the absorbed
laser pulse to $e^-e^+$ plasma can be seen in
Fig.~\ref{nvsx2} (line 1).

In conclusion we present the simulation results obtained 
by the developed two-dimensional hybrid PIC/MC model.  The
model allows us to simulate QED processes in laser plasmas.   
First we analyze the QED processes during interaction of
the relativistic electron beam with the intense laser pulse. It is shown
that the QED effects can be important even for moderately intense
laser pulses when the number of emitted photons by single electron is not
large.  Then avalanche-like cascade development in the field of two
colliding linearly polarized laser pulses is simulated. We show that overdense
($e^-e^+$) plasma is produced in the region where laser pulses
overlap.  A significant portion of the laser energy is absorbed as a
result of production and heating of the plasma. 
The anisotropic emission of hard photons is observed 
that can be used for development of bright radiation sources 
of gamma-quanta. The obtained results
demonstrate that QED effects can be observed in future laser facilities like ELI
\cite{eli} and HiPER \cite{hiper}, since the laser intensity ($I \approx 2\cdot 10^{24} \text{~W/cm}^2$) 
can be achievable in such laser systems.

We are grateful to A.~M.~Fedotov and N.~B.~Narozhny for fruitful
discussions.  This work has been supported by federal target program
"The scientific and scientific-pedagogical personnel of innovation in
Russia", by the Russian Foundation for Basic Research and by Dynasty
Foundation.

\end{document}